\documentclass[10pt,aps,prl,twocolumn,showpacs,superscriptaddress]{revtex4-1}  

\usepackage[modulo,switch]{lineno}
\modulolinenumbers[1]

\usepackage[utf8]{inputenc}
\usepackage{graphicx}
\usepackage{amsmath}
\usepackage{amssymb}
\usepackage{bm}        
\usepackage{txfonts}
\usepackage{multirow}
\usepackage{mathptmx}  
\usepackage{siunitx}
\usepackage{xargs}                      
\usepackage{microtype}

\usepackage[pdftex,dvipsnames]{xcolor}  
\usepackage[normalem]{ulem} 

\usepackage[%
  colorlinks=true,
  urlcolor=blue,
  linkcolor=blue,
  citecolor=blue
]{hyperref}
\usepackage{changes}
\definechangesauthor[name={Dmitry Bykov}, color=red]{DB}
\definechangesauthor[name={Lorenzo Dania}, color=blue]{LD}
\definechangesauthor[name={Tommaso Faorlin}, color=pink]{TF}
\definechangesauthor[name={Giovanni Cerchiari}, color=orange]{GC}

\begin{document}

\title{Back action suppression for levitated dipolar scatterers}

\author{Y.~Weiser}
\affiliation{Institut f\"ur Experimentalphysik, Universit\"at Innsbruck, Technikerstrasse~25, 6020~Innsbruck, Austria}
\author{T.~Faorlin}
\affiliation{Institut f\"ur Experimentalphysik, Universit\"at Innsbruck, Technikerstrasse~25, 6020~Innsbruck, Austria}
\author{L.~Panzl}
\affiliation{Institut f\"ur Experimentalphysik, Universit\"at Innsbruck, Technikerstrasse~25, 6020~Innsbruck, Austria}
\author{T.~Lafenthaler}
\affiliation{Institut f\"ur Experimentalphysik, Universit\"at Innsbruck, Technikerstrasse~25, 6020~Innsbruck, Austria}
\author{L.~Dania}
\affiliation{Institut f\"ur Experimentalphysik, Universit\"at Innsbruck, Technikerstrasse~25, 6020~Innsbruck, Austria}
\author{D.~S.~Bykov}
\affiliation{Institut f\"ur Experimentalphysik, Universit\"at Innsbruck, Technikerstrasse~25, 6020~Innsbruck, Austria}
\author{T.~Monz}
\affiliation{Institut f\"ur Experimentalphysik, Universit\"at Innsbruck, Technikerstrasse~25, 6020~Innsbruck, Austria}
\affiliation{AQT, Technikerstrasse~17, 6020~Innsbruck, Austria}
\author{R.~Blatt}
\affiliation{Institut f\"ur Experimentalphysik, Universit\"at Innsbruck, Technikerstrasse~25, 6020~Innsbruck, Austria}
\affiliation{Institut f\"ur Quantenoptik und Quanteninformation, \"Osterreichische Akademie der Wissenschaften, Technikerstrasse 21a, 6020 Innsbruck, Austria}
\author{G.~Cerchiari}
\email{giovanni.cerchiari@uibk.ac.at}
\altaffiliation{Corresponding author}
\affiliation{Institut f\"ur Experimentalphysik, Universit\"at Innsbruck, Technikerstrasse~25, 6020~Innsbruck, Austria}

\date{\today}

\begin{abstract}
Levitated dipolar scatterers exhibit exceptional performance as optomechanical systems for observing quantum mechanics at the mesoscopic scale. However, their tendency to scatter light in almost any direction poses experimental challenges, in particular limiting light collection efficiencies and, consequently, the information extractable from the system. In this article, we present a setup designed to enhance the information gleaned from optomechanical measurements by constraining the back action to a specific spatial direction. This approach facilitates achieving Heisenberg-limited detection at any given numerical aperture. The setup consists of a hollow hemispherical mirror that controls the light scattered by the dipolar emitter, particularly at high scattering angles, thereby focusing the obtained information. This mirror is compatible with existing setups commonly employed in levitated optomechanics, including confocal lenses and optical resonators.
\end{abstract}
\maketitle

\section{Introduction}
Levitated mesoscopic particles are noteworthy in the realm of optomechanical systems due to their high isolation from the environment. For instance, experiments with particles levitated in ultra-high vacuum (UHV) have recently demonstrated a damping rate of the center-of-mass (CoM) motion in the \qty{e-9}{\second^{-1}} regime, corresponding to a quality factor on the order of $10^{10}$~\cite{dania2023ultrahigh}. With the recent advances in quantum motion control~\cite{delic2019cavity,magrini2020,tebbenjohanns2021quantum,piotrowski2023simultaneous}, mesoscopic particles levitating in UHV provide an ideal platform for testing fundamental physics and sensing applications~\cite{Millen_2020,gonzalezballestero2021levitodynamics}. These optomechanical experiments hinge on the ability to observe and preserve motional quantum effects.  However, when light is involved, the scattering of photons induces back action stemming from the momentum transfer of the photons. Consequently, the ability of disturbance free measurement is constrained by the ratio between  the collection of scattered photons~\cite{Tebbenjohanns2019,cerchiari2021dipole} and the back action induced by the photon recoil ~\cite{Jain2016}. The optimum is achieved, when all the scattered photons can be detected. In this case, when a photon scatters and transfers momentum, it contributes to the overall measurement and minimizes the measurement uncertainty according to the Heisenberg limit~\cite{Tebbenjohanns2019}. However, levitated optomechanical masses scatter light in many directions, forcing experimenters to either compromise on solid angle coverage~\cite{delic2019cavity} or use deep parabolic mirrors~\cite{Salakhutdinov2020} to collect most of the emitted radiation. While the latter solution captures almost all scattered light, it limits the manipulation of the levitated object by dedicating most of the solid angle to light collection. The former solution leaks some scattered photons, making them unavailable for measurement. To reduce photon recoil noise, more complex approaches involving squeezed light~\cite{gonzalezballestero2023suppressing} or an auxiliary quantum system~\cite{Moller2017} may be necessary, requiring control of additional quantum degrees of freedom.

In this article, we suggest the use of a hollow hemispherical mirror to prevent the scattering of light and, consequently, to mitigate the photon back action. Furthermore, such a mirror could concentrate the information about the scatterer's position within a small solid angle, making it more readily accessible in the experiment. This solution has recently become technologically feasible with the development of high-precision hemispherical mirrors~\cite{Higginbottom2018} and apparatuses that can simultaneously accommodate such a hollow mirror and levitate a charged particle with electromagnetic fields~\cite{araneda2020}. The proposed mirror has the potential to selectively suppress scattering and its corresponding back action outside a small solid angle, effectively confining scattering to a narrow angular range. We provide a theoretical description of the process of scattering suppression, calculate the angular distribution of the remaining scattered photons, and discuss the limitations of the proposed techniques.

\section{Setup}

\begin{figure}
\includegraphics[width=\columnwidth]{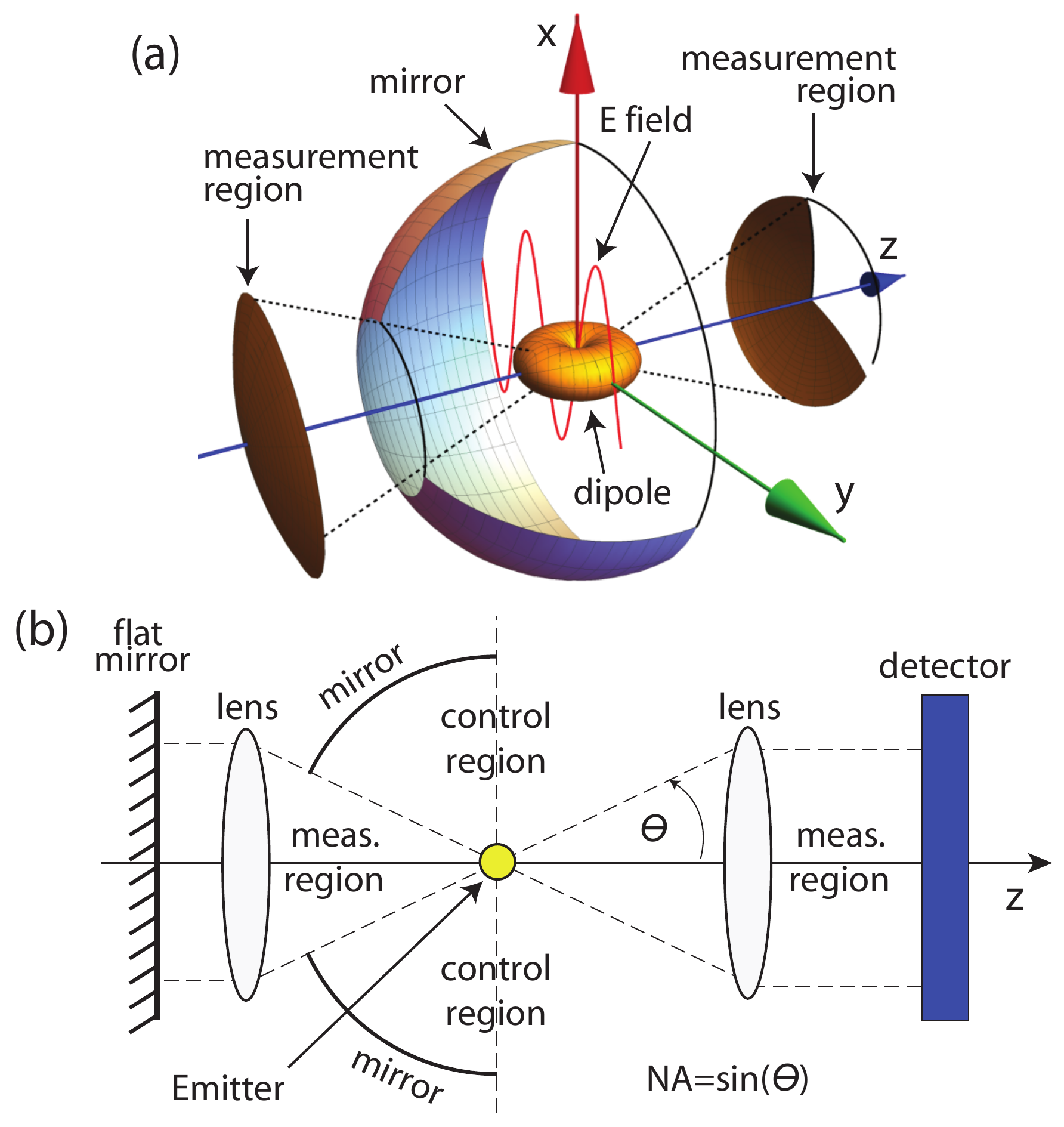}
\caption{(a) A dipolar scatterer is located in the center of curvature of a hollow hemispherical mirror and illuminated along the $y$-axis (``E-field''). The mirror controls the spontaneous emission of the scatterer in all directions of space except for a conical region around the $z$-axis (``measurement region''). (b) Schematic representation of how the hollow hemispherical mirror can be combined with existing optical setups such a confocal set of lenses. The depicted setup with a flat mirror and and flat detector can only reach the Heisenberg limit for the position measurement of a dipolar scatterer at limited numerical aperture ($\textrm{NA}=\sin{\left(\theta\right)}$) if the hemispherical mirror is present. The mirror applies suppression of scattering in the ``control region''. The portion of the control region opposite to the mirror is available to control the emitter.}
\label{fig:setup}
\end{figure}

Figure~\ref{fig:setup}(a) presents an experimental configuration in which a scatterer is located in the center of curvature of a hollow hemispherical mirror. The mirror divides the entire solid angle around the emitter into two distinct regions. The first one is called the ``control region'' and is the part of the solid angle in which influence on the scattering is exerted. The second one, named ``measurement region'', is the conical region around the optical axis that is not influenced by the mirror. The key idea discussed in this article is to use the mirror to suppress scattering in the control region so that it is only allowed in the measurement region. The purpose of the mirror is to reflect the radiation of the scatterer towards itself. In levitated optomechanical experiments, this system can be modeled in the far-field approximation. This can be seen by comparing the typical size of optical elements of about \SI{1}{\centi\meter}~\cite{Higginbottom2018} with the size of scatterers which are between \qty{100}{\nano\meter} and \qty{500}{\nano\meter}~\cite{dania2022} and the wavelength of the light radiation field which is in the range between \qty{500}{\nano\meter} to \qty{1500}{\nano\meter}. In the far-field approximation, light scattering is independent for each spatial direction. In this model, the hemispherical mirror controls the scattering of light only along the paths in which reflection occurs, leaving scattering in the other directions unaffected. This holds for linear dipolar scatterer. In the case of non-linear emitters, like atoms and ions, these two directions can not be treated independently \cite{Eschner2001}.

We consider scatterers that are described by a dipole distribution symmetric across the origin. Such a symmetry is common in the current state-of-the-art research that investigates optomechanical effects at the quantum level with levitated atoms~\cite{cerchiari2021one}, spheres~\cite{Delic2020}, rods~\cite{Martinetz_2021} or dumbbells~\cite{van2021sub}. The symmetry has expression $\rho\left(\bm{x}\right) = \rho\left(-\bm{x}\right)$, where the symbol $\rho$ refers to the dipole distribution and the vector $\bm{x}=(x,y,z)$ indicates the position in a Cartesian reference frame. Suppression of light scattering into the control region hinges on this symmetry. Therefore, we first discuss the implications in this configuration. In a later section we also consider the case where the scatterer has a non-symmetrical density distribution $\rho(\bm{x})$. In this article, $\rho\left(\bm{x}\right)$ is a real function that describes the volume density of a dipole moment inside the scatterer induced by a laser propagating with wavevector  $\bm{k}$ and having linear polarization $\hat{\bm{p}}$. 

In the far-field approximation, the differential power $d\tilde{P}$ scattered by the object in a generic direction $\hat{\bm{n}}=\left(\sin{\theta}\cos{\phi}, \sin{\theta}\sin{\phi}, \cos{\theta}\right)$ can be calculated using the Green's function $g\left(\hat{\bm{n}},\bm{x}\right)$ via the expression
\begin{equation}
    \label{eq:dP_full}
    \frac{d\tilde{P}}{d\Omega}= \frac{1}{p^2}\left\lvert\int\rho\left(\bm{x}\right) g\left(\hat{\bm{n}},\bm{x}\right)e^{i \bm{k}\cdot\bm{x}} d^3\bm{x}\right\rvert^2 \frac{dP}{d\Omega}\;,
\end{equation}
where
\begin{equation}
    \label{eq:dP_dipole}
    \frac{dP}{d\Omega} = p^2 \frac{c\lvert \bm{k}\rvert^4}{32 \pi^2 \epsilon_0}\left(1-\left(\hat{\bm{n}}\cdot \hat{\bm{p}}\right)^2\right)\;
\end{equation}
is the differential radiated power per unit solid angle of a point-like dipolar emitter having dipole moment $p$. In the expression, $c$ is the speed of light, $\bm{k}$ the wavevector of the illuminating light field, $\epsilon_0$ the electric permittivity of vacuum and the term $\exp{\left(i \bm{k}\cdot\bm{x}\right)}$ is the phase difference introduced by the propagating light beam. This expression is derived in detail in Appendix A. In the case of a non point-like object, the total dipole moment can be calculated by integrating the distribution $\rho(\bm{x})$ in space: $p=\int \rho(\bm{x}) d^3\bm{x}$.

In the measurement region, the function $g$ corresponds to the emission of a free space emitter, which is~\cite{hulst1981light}
\begin{equation}
    g_m\left(\hat{\bm{n}},\bm{x}\right) = e^{i\lvert \bm{k}\rvert\hat{\bm{n}}\cdot\bm{x}} \; .
\end{equation}
as given by the far-field approximation of the Huygens-Fresnel integral on a sphere of unitary radius. In the control region, where the hemispherical mirror suppresses light scattering, the Green's function is
\begin{equation}\label{eq:green_mirror}
    g_c\left(\hat{\bm{n}},\bm{x}\right)=2 \sin{\left(\lvert \bm{k}\rvert\hat{\bm{n}}\cdot\bm{x}\right)}  \; .
\end{equation}
No light can escape from the side where the mirror is installed, so the integral in Eq.~\eqref{eq:dP_full} corresponding to the control region is only relevant on the opposite side of this boundary condition. The mirror can be used to control the scattering rate via the sinusoidal function. For instance, if the object's extension in the plane orthogonal to $\bm{k}$ is small compared to the wavelength, the object will appear dark in the control region, because the integrand of Eq.~\eqref{eq:dP_full} is the product between a symmetric function $\rho\left(\bm{x}\right)$ and an antisymmetric function $\sin{\left(\lvert \bm{k}\rvert\hat{\bm{n}}\cdot\bm{x}\right)}$.

This result can be expanded to all directions of space in order to tackle more practical problems. By limiting the size of the object to be smaller than the wavelength of the illuminating light in every direction, the expansion of the function $e^{i \bm{k}\cdot\bm{x}} g\left(\hat{\bm{n}},\bm{x}\right)$ to the leading order in the normalized distance $\lvert \bm{x}\rvert/\lambda$ becomes reasonable. Under this approximation, we find
\begin{align}
    \label{eq:green_free_leading}
    e^{i \bm{k}\cdot\bm{x}} g_m\left(\hat{\bm{n}},\bm{x}\right) &\approx 1,  \\
    \label{eq:green_mirror_leading}
    e^{i \bm{k}\cdot\bm{x}} g_c\left(\hat{\bm{n}},\bm{x}\right) &\approx 2 \lvert \bm{k}\rvert \left(\hat{\bm{n}}\cdot\bm{x}\right). \; 
\end{align}
From the last equations, one can observe that $g_c$ is an antisymmetric function up to first order and thus, for a small symmetric object, the integral of Eq.~\eqref{eq:dP_full} is zero even when the object extends in the direction of the illuminating beam. In contrast, the function $g_m$ contains a constant term that dominates the scattering rate. It must be noted that for objects with a significant extension compared to the wavelength the suppression is only approximate and we quantify the effect of a finite size radius for a scattering levitated sphere in Appendix B.

\begin{figure}
\includegraphics[width=\columnwidth]{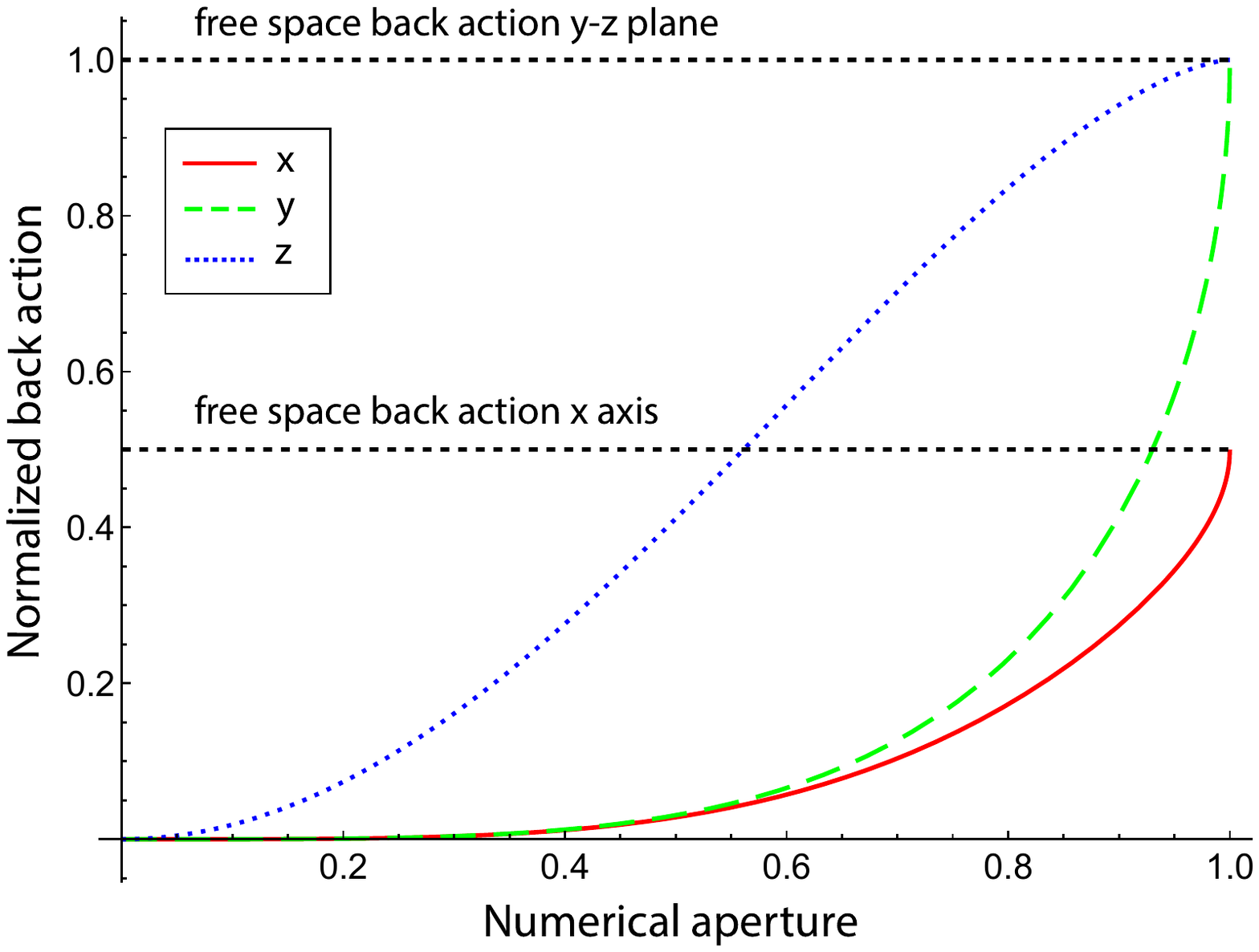}
\caption{The back action as a function of the numerical aperture of the measurement region, as given in Eq.~\eqref{eq:sax} -~\eqref{eq:saz}. Hereby, the graphs were normalized to the maximal back action $S_{ba}^z=2 \hbar \lvert \bm{k}\rvert P_{dip}/(10 \pi c)$ of the free space emitter. The continuous-red, the dashed-green and the dotted-blue line describe the normalized back action in $x$, $y$ and $z$ direction respectively. As shown in Fig.~\ref{fig:setup} the measurement region is oriented along the $z$ axis. Thus, the back action is mainly affecting the motion along the $z$ direction, whereas the back action in the other directions is well suppressed even for higher NAs of the measurement region. Since the polarization of the incident light is parallel to the $x$ axis the free space back action in this direction is smaller compared to the one in $y$ and $z$ direction.}
\label{fig:back_action_cm}
\end{figure}

\section{Back action suppression}

The selective suppression of the emission introduced by the mirror results in a selective suppression of the corresponding optomechanical back action. The differential back action $ds_{ba}$ for radiation scattered in the differential solid angle $d\Omega$ under the direction $\hat{\bm{n}}$ originates from the power spectral density fluctuation of the radiation pressure force and has expression~\cite{Tebbenjohanns2019, cerchiari2021dipole}
\begin{equation}
    \label{eq:differential_back_action}
    \frac{ds_{ba}}{d \Omega} = \frac{\hbar \lvert \bm{k}\rvert}{2 \pi c} \left(\hat{\bm{n}}\cdot \bm{x}_0\right)^2 \frac{dP}{d \Omega}\;,
\end{equation}
where $\bm{x}_0$ is the displacement of the object's center of mass from the origin. With the hemispherical mirror, scattering only occurs in the measurement region. Therefore, the total back action $S_{ba}$ calculated for a displacement adhering to the condition $\lvert\bm{x}_0\rvert=1$ depends on the angle of aperture $\theta_h$ of the hole inside the mirror via the following expression
\begin{equation}
    \label{eq:total_back_action}
    S_{ba}\left(\hat{\bm{x}}_0, \theta_h\right) = 2\int_{0}^{\theta_h}d\theta \sin{\left(\theta\right)}\int_{0}^{2\pi}d\phi \frac{ds_{ba}}{d\Omega} \;.
\end{equation}
The factor of two in front accounts for the region of the hole and on the opposite side. We align the polarization of the light beam along the $x$ axis ($\hat{\bm{p}}=\hat{\bm{x}}$) and normalize our result to the maximal back action of the free space emitter that is $S_{ba}^z = S_{ba}\left(\hat{\bm{z}}, \pi/2\right) =  2 \hbar \lvert \bm{k}\rvert P_{dip}/(10 \pi c)$ experienced by the scatterer in the plane orthogonal to the light polarization. Figure~\ref{fig:back_action_cm} presents the normalized back action as a function of the numerical aperture of the measurement region. The full analytical formulas of the presented curves can be found in Appendix~C. Here, we report the expressions of the curves expanded to fourth order as a function of the aperture angle $\theta$ near the condition $\textrm{NA}=\sin{\left(\theta\right)}=0$. The expressions are
\begin{align}
    \frac{\bm{S}_{ba}\left(\hat{\bm{x}}, \theta\right)}{\bm{S}_{ba}^z} &= \frac{15}{16}\theta^4 \;, \label{eq:sax}\\
    \frac{\bm{S}_{ba}\left(\hat{\bm{y}}, \theta\right)}{\bm{S}_{ba}^z} &= \frac{15}{32}\theta^4 \;, \label{eq:say}\\
    \frac{\bm{S}_{ba}\left(\hat{\bm{z}}, \theta\right)}{\bm{S}_{ba}^z} &= \frac{15}{8} \theta^2 - \frac{25}{16} \theta^4 \label{eq:saz}\;.
\end{align}
We see that the leading orders in the expansions indicate that the back action in the $x$-$y$ plane compared to the back action in the $z$ direction is suppressed because of the different dependency on the angle $\theta$. As an example, we select a numerical aperture of $\textrm{NA}=0.4$, which has proven to be a feasible choice for measurements of the position operator at the quantum level~\cite{cerchiari2021one}. At this $\textrm{NA}$, the back action along the mirror's axis is about $10^3$ times larger than in the orthogonal plane.

For this calculation, we consider the hybrid system consisting of the hollow mirror and a self-homodyne detection apparatus~\cite{Bushev2006, dania2022} that should detect the object's motion by utilizing the measurement region. This configuration is depicted in Fig.~\ref{fig:setup}(b), which presents how the two techniques can be combined. In the self-homodyne method, a pair of confocal lenses collimate the scattered light onto a flat mirror and a flat detector. The flat mirror is adjusted to generate a position-dependent modulation of the power emitted by the scatterer, which can be used to analyze the scatterer-mirror distance from the light intensity recorded by the detector. Without the hemispherical mirror, this position-detection method is predicted to reach the Heisenberg limit in the limit in which the lenses have unitary NA. For such a setup, the minimal measurement imprecision $s$ per unit solid angle is~\cite{cerchiari2021dipole}
\begin{equation}
    \label{eq:differential_imprecision}
    s = \frac{\hbar c}{32\pi \lvert\bm{k}\rvert} \frac{1}{\left(\hat{\bm{n}} \cdot \bm{x}_0\right)^2 dP/d\Omega} \;.
\end{equation}
This expression corresponds to the variance of the position measurement for a differential detector. Since the imprecision is a variance, the total minimal imprecision $S_{imp}$ for a full hemispherical detector is obtained by inverse weighting, according to the expression $S_{imp}=\left(\int s^{-1} d\Omega\right)^{-1}$. Here, in contrast to previous literature~\cite{cerchiari2021dipole}, the angular integral follows the same domain as specified in Eq.~\ref{eq:total_back_action} because no detection is meaningful in the control region, which is equivalent to assuming that the detector should cover only the measurement region and not the entire hemisphere. We see that the angular dependence of the back action (Eq.~\ref{eq:differential_back_action}) and the imprecision (Eq.~\ref{eq:differential_imprecision}) are the inverse of each other up to a constant. For this reason, if we compute the product of the total measurement imprecision and the total back action of the hybrid system, we find a constant. The constant is equal to $\hbar^2/(32\pi^2)$ and corresponds to the maximal achievable product between measurement imprecision and back action set by the Heisenberg uncertainty relation~\cite{Clerk}. Therefore, we see that the hollow hemispherical mirror enables the measurement of position at the Heisenberg limit at limited NA.

Combining the measurement at the Heisenberg limit with the suppressed back action in the $x$-$y$ plane, we can conclude that the hollow mirror can be used to reduce the theoretical analysis and control of a levitated scatterer to a monodimensional problem along the mirror's optical axis. This is justified by the fact that, for small aperture angles of the mirror's hole, the setup grants Heisenberg-limited detection of the center of mass position along the only axis influenced by the measurement back action. As an additional benefit, we note that, by leveraging this property, experimenters can work in the paraxial approximation, making it possible to ignore the modulation to the emitted radiation field typical of dipolar patterns, which simplifies the theoretical description of the levitated system.

\section{Asymmetries in the spatial distribution}\label{sec:asy}

If the induced-dipole distribution $\rho\left(\bm{x}\right)$ is not symmetric with respect to the hemispherical mirror's center of curvature, the suppression of the emission is not complete. The purpose of this section is to discuss this potential limitation to show that the scattered field caused by asymmetries is not affecting the momentum in the same way as for a free-space emitter. The scattered light can be analyzed to perform measurements of different motional quantities such as the variance of the center of mass and the libration angles. This discussion is only relevant for the control region, therefore, we present our calculation for a full hemispherical mirror, i.e. if the measurement region of Fig.\ref{fig:setup}(a) is negligibly small. We briefly comment on the implications for measurements with a hollow hemispherical mirror.

\begin{figure*}
\includegraphics[width=\textwidth]{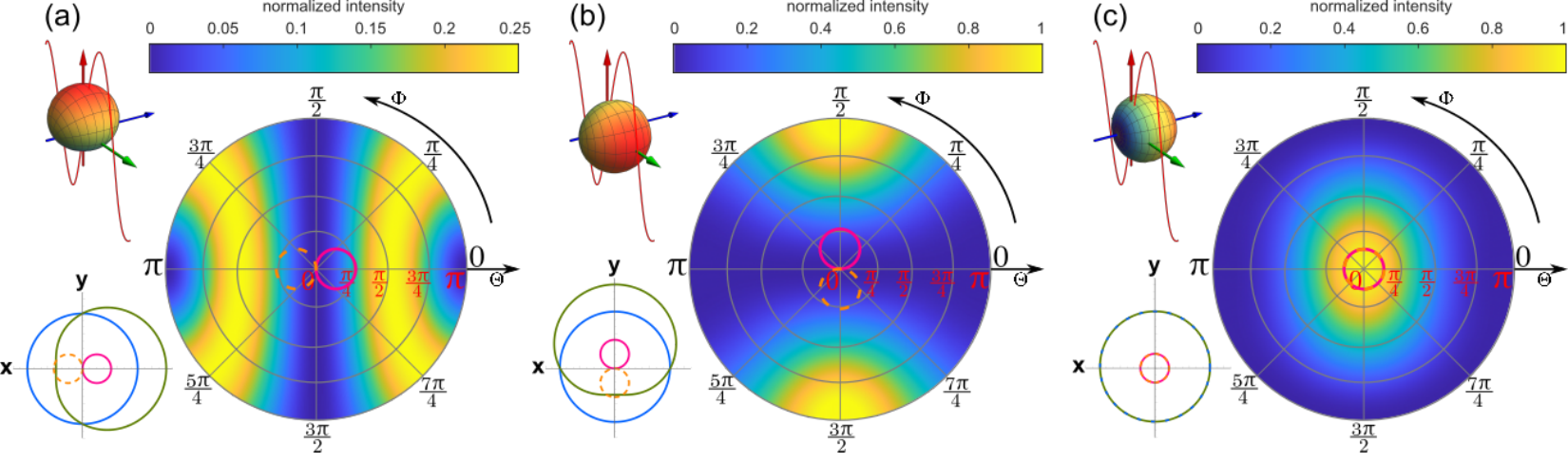}
\caption{The figure shows the light power scattered at different polar ($\theta$) and azimuthal ($\phi$) angles by an asymmetric emitter located in the center of curvature of a full hemispherical mirror. We used the map $(x,y,z)=r(\sin{\theta}\cos{\phi},\sin{\theta}\sin{\phi},\cos{\theta})$. In the subfigures, the asymmetry 
is oriented along the $x$ (a), $y$ (b), and $z$ (c) direction. A three-dimensional model of the emitter is presented in the upper left corner of each subfigure with the same perspective of Fig.~\ref{fig:setup}(a). The sinusoidal red line in the three-dimensional model indicates the polarization of the light along the $x$-axis ($\theta=\pi/2$, $\phi=0$). At the lower-left corner of each subfigure, a bi-dimensional section of the emitter in the $x$-$y$ plane presents the emitter itself (green), its symmetrical part (blue), and its antisymmetrical part (pink) parts as well as the reflection (orange) that modifies the Green's function. In the middle of each angular plot, the antisymmetrical part and its back reflection are drawn in pink (solid) and orange (dashed) respectively. We used the asymmetry parameter $c_0/C_{00}=0.3$ (see Eqs.~\ref{eq:multipole_expansion}, \ref{eq:multipole_expansion_vector} in the main text) and we normalized the colorbars to the maximal power which is obtained at $\theta=0$ in subplot (c).}
\label{fig:asym}
\end{figure*}

One of the possible perturbations that causes asymmetries are displacements of the symmetric object from the origin. For levitated optomechanical experiments operating in the quantum regime, displacements $\bm{x}_0$ are small compared to the light wavelength and the spatial distribution. For example, a levitated SiO$_2$ particle composed of $\sim10^9$ atoms has a radius of $\sim100$~\si{\nano\meter} and has a single quantum displacement of the center of mass of \SI{10}{\pico\meter} to \SI{100}{\pico\meter}. Therefore, the leading contribution to the scattered field can be calculated by expanding the density $\rho$ around the origin as a function of the displacement arriving from Eqs.~\ref{eq:dP_full} and \ref{eq:green_mirror_leading} in a few steps to
\begin{equation}
    \label{eq:variance}
    \frac{d \tilde{P}}{d \Omega}= 4 \lvert \bm{k}\rvert^2 \left\lvert \left( \int \nabla \rho \; \left(\hat{\bm{n}}\cdot\bm{x}\right) d^3\bm{x}\right) \cdot \bm{x}_0\;\right\rvert^2 \frac{d P}{d \Omega}.
\end{equation}
This term is not null because the gradient $\nabla \rho$ of a symmetric distribution $\rho$ is antisymmetric. The vector $\bm{\gamma} = \int \nabla \rho \; \left(\hat{\bm{n}}\cdot\bm{x}\right) d^3\bm{x}$ is a geometric factor that depends on the shape of the object. For example, for a levitated sphere of radius $r_0$, which has a total dipole moment of $p=\frac{4\pi}{3}\rho r_0^3$ it is equal to $\bm{\gamma}=\frac{p}{2} \hat{\bm{n}}$. Using this result in Eq.~\ref{eq:variance} we obtain
\begin{equation}
    d\tilde{P} = \lvert \bm{k}\rvert^2 \lvert\hat{\bm{n}}\cdot\bm{x}_0\rvert^2 dP \;.
\end{equation}
We see that the scattered power increases quadratically as a function of the displacement and, therefore, implements the measurement of the variance operator, which has been proposed as a possible means to measure or prepare quantum states of motion~\cite{RomeroIsart2011,Brawley2016,Aspelmeyer2014}. 

A second contribution to asymmetries is given by intrinsic inhomogeneities of the object's dipole distribution $\rho$ with respect to its mass density. For sub-wavelength-sized objects, we model the dominant contribution of these asymmetries by expanding the function $\rho\left(\bm{x}\right)$ with the help of spherical harmonics $Y_{lm}$ to the first order arriving at
\begin{equation}
    \label{eq:multipole_expansion}
    \rho\left(\theta,\phi\right) = \frac{p}{2\sqrt{\pi}}  \left(C_{00}Y_{00}+C_{1-1}Y_{1-1}+C_{10}Y_{10}+C_{11}Y_{11}\right)\;.
\end{equation}
The expression is normalized so that the total dipole moment can be calculated as $p = \int \rho\left(\theta,\phi\right) d\Omega$. The constant $C_{00}$ scales the overall size of the object and has the dimension of length. Substituting this distribution in Eq.~\ref{eq:dP_full}, and using Eq.~\ref{eq:green_mirror_leading}, we find that the scattered power is 
\begin{equation}
    \label{eq:orientation}
    \frac{d\tilde{P}}{d\Omega}= \frac{4 \lvert\bm{k}\rvert^2}{p^2}\left\lvert\int\rho\left(\theta,\phi\right) \left(\hat{\bm{n}}\cdot\hat{\bm{x}}\right)d\Omega\right\rvert^2 \frac{dP}{d\Omega}\;.
\end{equation}
In the presence of the mirror, the contribution to scattering arises only from the terms $Y_{1m}$ because $Y_{00}$ is symmetric. We further select the vector $\bm{C}=\left(C_{1-1}, C_{10}, C_{11}\right)$ to describe a generic asymmetry oriented towards the polar angle $\theta_1$ and the azimuthal angle $\phi_1$ with the expression
\begin{equation}
    \label{eq:multipole_expansion_vector}
    \bm{C}=c_0\left(\sin{\theta_1}\frac{e^{i\phi_1}}{\sqrt{2}},\cos{\theta_1},-\sin{\theta_1}\frac{e^{-i\phi_1}}{\sqrt{2}}\right)\;,
\end{equation}
where $c_0$ is a constant determining the size of the asymmetry and has the dimension of length.
The scattered power resulting from this asymmetry is obtained using Eq.~\ref{eq:orientation} to find
\begin{equation}
    d\tilde{P}=\frac{\lvert\bm{k}\rvert^2 c^2_0}{3} \left(c_{\theta} c_{\theta_1} + 
 s_{\theta} s_{\theta_1} \cos{\left(\phi - \phi_1\right)}\right)^2 dP\;,
\end{equation}
where $c_\alpha = \cos{(\alpha)}$ and $s_\beta = \sin{(\beta)}$.
This is a further shaping of the emission pattern that can be used to deduce the orientation of the object. An example of the power scattered by such a first-order asymmetry is given in Fig.~\ref{fig:asym}, where the orientation of the object is shown, together with the respective angular power distribution and the decomposition of the object in its symmetric and asymmetric parts. In real experiments, intrinsic asymmetries of the spatial distribution tend to align along the polarization of the electric field of the illuminating light~\cite{Hoang2016}, which corresponds to the conditions $\phi_1=0$ and $\theta_1=\pi/2$ and, so, to the scattering pattern presented in Fig.~\ref{fig:asym}(a). This is the configuration with minimal excess scattering induced by the asymmetry: it is two times lower compared to the maximum configurations in which the scatterer is oriented in the $y$-$z$ plane (Fig.~\ref{fig:asym}(b) and \ref{fig:asym}(c)). Besides, for $x$ oriented asymmetry, we note that little light, and thus little information, regarding the asymmetry's orientation is directed at low polar angle $\theta$. If we allow the mirror to be hollow and assume a measurement region having a numerical aperture of $\textrm{NA}=0.4$, we calculate that only 2\% of the total scattered power that characterizes the pattern presented in Fig.~\ref{fig:asym}(a) is lost in the direction of the hole. This fact suggests that by employing a pixel detector on the opposite side of the mirror to separate the different angular regions, the measurements of the center of mass' position and the asymmetry's orientation could be performed simultaneously. In contrast to state-of-the-art methods that rely on polarization fluctuations, this technique for orientation detection has the potential advantage to rely only on modulation of the detected intensity. However, it is not applicable to objects if they have radial symmetry with respect to the center of mass such as symmetric dumbbell~\cite{Ahn2018} or homogeneous nanorods~\cite{Kuhn2017} that are currently the main focus of modern research in the field for the topic of orientation detection. Quantifying the performance of such a measurement of the libration angle and propose suitable levitated scatterers for its applicability is beyond the scope of this article and we defer it to future works.

Finally, we conclude this section underlining that both the scattered power $d\tilde{P}$ determined by displacements or by intrinsic asymmetries are suppressed if compared to the scattered power $dP$ by a free-space emitter because of the pre-factors $\lvert \bm{k}\rvert^2 \lvert \bm{x}_0\rvert^2$ and $\lvert \bm{k}\rvert^2 c_0^2$. These factors emerge from the fact that the object is located in the minimum of an interference fringe (see Eq.~\ref{eq:green_mirror}) which is the reason why the scattered light is suppressed.

\section{Conclusion}
We have shown that a hollow hemispherical mirror can be used to control the photon back action acting on a scatterer placed at the center of the curvature of the mirror by constraining the scattering to a selected solid angle, while, at the same time, enabling particle position measurement at the Heisenberg limit. This allows for reducing the treatment and the behavior of the optomechanical system to a single direction of space along the direction of the mirror's hole. This setup enables manipulating the dipolar scatterer without compromising on the solid angle of detection because, for example, it enables electromagnetic trapping with a Paul trap~\cite{araneda2020} by locating the trap's electrodes in the solid angle facing the mirror where scattering is suppressed. 

The limitations of this setup in suppressing light scattering stem from asymmetries in the spatial distribution of the levitated object such as displacements of the object with respect to the mirror's center of curvature or intrinsic asymmetries of the object's shape. Such asymmetries lead to light scattering which can be detected to measure the variance operator of the center of mass motion as well as to reconstruct the orientation of the object's asymmetry. These complementary measurements do not affect the motional degrees of freedom in the same way as for a free space emitter. This fact may enable the detection of objects' shape and orientation without perturbing the center of mass momentum.

Besides allowing for a more precise position measurement, the reduction of back action can also be beneficial for state detection via photon emission. The heating of ions due to multiple scattering events for high fidelity read out is already a limitation in some systems \cite{Krutyanskiy2023}. Since the read out of qubit states in trapped ion systems is often done via fluorescence detection, an increase in detection efficiency due to a spatially tailored emission pattern, like for the hole in a hemispherical mirror, can be expected. This can lower the time needed for cooling of the ion after state detection.

The calculations described in this article are available as scripts written for Wolfram Mathematica at the repository~\cite{Cerchiari2024}.

\hfill \break
\textit{Acknowledgements.} This work was supported by the Austrian Science Fund (FWF) project number: P 36233-N (SONATINA).
This work was also supported by the Institut für Quanteninformation GmbH. 

\bibliography{bibliography}

\begin{thebibliography}{33}
\expandafter\ifx\csname natexlab\endcsname\relax\def\natexlab#1{#1}\fi
\expandafter\ifx\csname bibnamefont\endcsname\relax
  \def\bibnamefont#1{#1}\fi
\expandafter\ifx\csname bibfnamefont\endcsname\relax
  \def\bibfnamefont#1{#1}\fi
\expandafter\ifx\csname citenamefont\endcsname\relax
  \def\citenamefont#1{#1}\fi
\expandafter\ifx\csname url\endcsname\relax
  \def\url#1{\texttt{#1}}\fi
\expandafter\ifx\csname urlprefix\endcsname\relax\def\urlprefix{URL }\fi
\providecommand{\bibinfo}[2]{#2}
\providecommand{\eprint}[2][]{\url{#2}}

\bibitem[{\citenamefont{Dania and Bykov~\textit{et
  al.}}(2023)}]{dania2023ultrahigh}
\bibinfo{author}{\bibfnamefont{L.}~\bibnamefont{Dania}} \bibnamefont{and}
  \bibinfo{author}{\bibfnamefont{D.~S.} \bibnamefont{Bykov~\textit{et al.}}},
  \bibinfo{journal}{arXiv:2304.02408 [quant-ph]}  (\bibinfo{year}{2023}),
  \urlprefix\url{https://arxiv.org/abs/2304.02408}.

\bibitem[{\citenamefont{Deli\ifmmode~\acute{c}\else \'{c}\fi{}
  et~al.}(2019)\citenamefont{Deli\ifmmode~\acute{c}\else \'{c}\fi{},
  Reisenbauer, Grass, Kiesel, Vuleti\ifmmode~\acute{c}\else \'{c}\fi{}, and
  Aspelmeyer}}]{delic2019cavity}
\bibinfo{author}{\bibfnamefont{U.~c.~v.}
  \bibnamefont{Deli\ifmmode~\acute{c}\else \'{c}\fi{}}},
  \bibinfo{author}{\bibfnamefont{M.}~\bibnamefont{Reisenbauer}},
  \bibinfo{author}{\bibfnamefont{D.}~\bibnamefont{Grass}},
  \bibinfo{author}{\bibfnamefont{N.}~\bibnamefont{Kiesel}},
  \bibinfo{author}{\bibfnamefont{V.}~\bibnamefont{Vuleti\ifmmode~\acute{c}\else
  \'{c}\fi{}}}, \bibnamefont{and}
  \bibinfo{author}{\bibfnamefont{M.}~\bibnamefont{Aspelmeyer}},
  \bibinfo{journal}{Phys. Rev. Lett.} \textbf{\bibinfo{volume}{122}},
  \bibinfo{pages}{123602} (\bibinfo{year}{2019}),
  \urlprefix\url{https://doi.org/10.1103/PhysRevLett.122.123602}.

\bibitem[{\citenamefont{Magrini et~al.}(2021)\citenamefont{Magrini, Rosenzweig,
  Bach, Deutschmann-Olek, Hofer, Hong, Kiesel, Kugi, and
  Aspelmeyer}}]{magrini2020}
\bibinfo{author}{\bibfnamefont{L.}~\bibnamefont{Magrini}},
  \bibinfo{author}{\bibfnamefont{P.}~\bibnamefont{Rosenzweig}},
  \bibinfo{author}{\bibfnamefont{C.}~\bibnamefont{Bach}},
  \bibinfo{author}{\bibfnamefont{A.}~\bibnamefont{Deutschmann-Olek}},
  \bibinfo{author}{\bibfnamefont{S.~G.} \bibnamefont{Hofer}},
  \bibinfo{author}{\bibfnamefont{S.}~\bibnamefont{Hong}},
  \bibinfo{author}{\bibfnamefont{N.}~\bibnamefont{Kiesel}},
  \bibinfo{author}{\bibfnamefont{A.}~\bibnamefont{Kugi}}, \bibnamefont{and}
  \bibinfo{author}{\bibfnamefont{M.}~\bibnamefont{Aspelmeyer}},
  \bibinfo{journal}{Nature} \textbf{\bibinfo{volume}{595}},
  \bibinfo{pages}{373} (\bibinfo{year}{2021}), ISSN \bibinfo{issn}{1476-4687},
  \urlprefix\url{https://doi.org/10.1038/s41586-021-03602-3}.

\bibitem[{\citenamefont{Tebbenjohanns et~al.}(2021)\citenamefont{Tebbenjohanns,
  Mattana, Rossi, Frimmer, and Novotny}}]{tebbenjohanns2021quantum}
\bibinfo{author}{\bibfnamefont{F.}~\bibnamefont{Tebbenjohanns}},
  \bibinfo{author}{\bibfnamefont{M.~L.} \bibnamefont{Mattana}},
  \bibinfo{author}{\bibfnamefont{M.}~\bibnamefont{Rossi}},
  \bibinfo{author}{\bibfnamefont{M.}~\bibnamefont{Frimmer}}, \bibnamefont{and}
  \bibinfo{author}{\bibfnamefont{L.}~\bibnamefont{Novotny}},
  \bibinfo{journal}{Nature} \textbf{\bibinfo{volume}{595}},
  \bibinfo{pages}{378} (\bibinfo{year}{2021}), ISSN \bibinfo{issn}{1476-4687},
  \urlprefix\url{https://doi.org/10.1038/s41586-021-03617-w}.

\bibitem[{\citenamefont{Piotrowski et~al.}(2023)\citenamefont{Piotrowski,
  Windey, Vijayan, Gonzalez-Ballestero, de~los R{\'\i}os~Sommer, Meyer,
  Quidant, Romero-Isart, Reimann, and Novotny}}]{piotrowski2023simultaneous}
\bibinfo{author}{\bibfnamefont{J.}~\bibnamefont{Piotrowski}},
  \bibinfo{author}{\bibfnamefont{D.}~\bibnamefont{Windey}},
  \bibinfo{author}{\bibfnamefont{J.}~\bibnamefont{Vijayan}},
  \bibinfo{author}{\bibfnamefont{C.}~\bibnamefont{Gonzalez-Ballestero}},
  \bibinfo{author}{\bibfnamefont{A.}~\bibnamefont{de~los R{\'\i}os~Sommer}},
  \bibinfo{author}{\bibfnamefont{N.}~\bibnamefont{Meyer}},
  \bibinfo{author}{\bibfnamefont{R.}~\bibnamefont{Quidant}},
  \bibinfo{author}{\bibfnamefont{O.}~\bibnamefont{Romero-Isart}},
  \bibinfo{author}{\bibfnamefont{R.}~\bibnamefont{Reimann}}, \bibnamefont{and}
  \bibinfo{author}{\bibfnamefont{L.}~\bibnamefont{Novotny}},
  \bibinfo{journal}{Nat. Phys.}  (\bibinfo{year}{2023}),
  \urlprefix\url{https://doi.org/10.1038/s41567-023-01956-1}.

\bibitem[{\citenamefont{Millen et~al.}(2020)\citenamefont{Millen, Monteiro,
  Pettit, and Vamivakas}}]{Millen_2020}
\bibinfo{author}{\bibfnamefont{J.}~\bibnamefont{Millen}},
  \bibinfo{author}{\bibfnamefont{T.~S.} \bibnamefont{Monteiro}},
  \bibinfo{author}{\bibfnamefont{R.}~\bibnamefont{Pettit}}, \bibnamefont{and}
  \bibinfo{author}{\bibfnamefont{A.~N.} \bibnamefont{Vamivakas}},
  \bibinfo{journal}{Rep. Prog. Phys.} \textbf{\bibinfo{volume}{83}},
  \bibinfo{pages}{026401} (\bibinfo{year}{2020}),
  \urlprefix\url{https://doi.org/10.1088/1361-6633/ab6100}.

\bibitem[{\citenamefont{Gonzalez-Ballestero
  et~al.}(2021)\citenamefont{Gonzalez-Ballestero, Aspelmeyer, Novotny, Quidant,
  and Romero-Isart}}]{gonzalezballestero2021levitodynamics}
\bibinfo{author}{\bibfnamefont{C.}~\bibnamefont{Gonzalez-Ballestero}},
  \bibinfo{author}{\bibfnamefont{M.}~\bibnamefont{Aspelmeyer}},
  \bibinfo{author}{\bibfnamefont{L.}~\bibnamefont{Novotny}},
  \bibinfo{author}{\bibfnamefont{R.}~\bibnamefont{Quidant}}, \bibnamefont{and}
  \bibinfo{author}{\bibfnamefont{O.}~\bibnamefont{Romero-Isart}},
  \bibinfo{journal}{Science} \textbf{\bibinfo{volume}{374}},
  \bibinfo{pages}{168} (\bibinfo{year}{2021}),
  \urlprefix\url{https://www.science.org/doi/abs/10.1126/science.abg3027}.

\bibitem[{\citenamefont{Tebbenjohanns et~al.}(2019)\citenamefont{Tebbenjohanns,
  Frimmer, and Novotny}}]{Tebbenjohanns2019}
\bibinfo{author}{\bibfnamefont{F.}~\bibnamefont{Tebbenjohanns}},
  \bibinfo{author}{\bibfnamefont{M.}~\bibnamefont{Frimmer}}, \bibnamefont{and}
  \bibinfo{author}{\bibfnamefont{L.}~\bibnamefont{Novotny}},
  \bibinfo{journal}{Phys. Rev. A} \textbf{\bibinfo{volume}{100}},
  \bibinfo{pages}{043821} (\bibinfo{year}{2019}),
  \urlprefix\url{https://doi.org/10.1103/PhysRevA.100.043821}.

\bibitem[{\citenamefont{Cerchiari
  et~al.}(2021{\natexlab{a}})\citenamefont{Cerchiari, Dania, Bykov, Blatt, and
  Northup}}]{cerchiari2021dipole}
\bibinfo{author}{\bibfnamefont{G.}~\bibnamefont{Cerchiari}},
  \bibinfo{author}{\bibfnamefont{L.}~\bibnamefont{Dania}},
  \bibinfo{author}{\bibfnamefont{D.~S.} \bibnamefont{Bykov}},
  \bibinfo{author}{\bibfnamefont{R.}~\bibnamefont{Blatt}}, \bibnamefont{and}
  \bibinfo{author}{\bibfnamefont{T.~E.} \bibnamefont{Northup}},
  \bibinfo{journal}{Phys. Rev. A} \textbf{\bibinfo{volume}{104}},
  \bibinfo{pages}{053523} (\bibinfo{year}{2021}{\natexlab{a}}),
  \urlprefix\url{https://link.aps.org/doi/10.1103/PhysRevA.104.053523}.

\bibitem[{\citenamefont{Jain et~al.}(2016)\citenamefont{Jain, Gieseler, Moritz,
  Dellago, Quidant, and Novotny}}]{Jain2016}
\bibinfo{author}{\bibfnamefont{V.}~\bibnamefont{Jain}},
  \bibinfo{author}{\bibfnamefont{J.}~\bibnamefont{Gieseler}},
  \bibinfo{author}{\bibfnamefont{C.}~\bibnamefont{Moritz}},
  \bibinfo{author}{\bibfnamefont{C.}~\bibnamefont{Dellago}},
  \bibinfo{author}{\bibfnamefont{R.}~\bibnamefont{Quidant}}, \bibnamefont{and}
  \bibinfo{author}{\bibfnamefont{L.}~\bibnamefont{Novotny}},
  \bibinfo{journal}{Phys. Rev. Lett.} \textbf{\bibinfo{volume}{116}},
  \bibinfo{pages}{243601} (\bibinfo{year}{2016}),
  \urlprefix\url{https://doi.org/10.1103/PhysRevLett.116.243601}.

\bibitem[{\citenamefont{Salakhutdinov}(2020)}]{Salakhutdinov2020}
\bibinfo{author}{\bibfnamefont{V.}~\bibnamefont{Salakhutdinov}},
  \bibinfo{type}{doctoral thesis},
  \bibinfo{school}{Friedrich-Alexander-Universit{\"a}t Erlangen-N{\"u}rnberg
  (FAU)} (\bibinfo{year}{2020}),
  \urlprefix\url{urn:nbn:de:bvb:29-opus4-137383}.

\bibitem[{\citenamefont{Gonzalez-Ballestero
  et~al.}(2023)\citenamefont{Gonzalez-Ballestero, Zieli\ifmmode~\acute{n}\else
  \'{n}\fi{}ska, Rossi, Militaru, Frimmer, Novotny, Maurer, and
  Romero-Isart}}]{gonzalezballestero2023suppressing}
\bibinfo{author}{\bibfnamefont{C.}~\bibnamefont{Gonzalez-Ballestero}},
  \bibinfo{author}{\bibfnamefont{J.}~\bibnamefont{Zieli\ifmmode~\acute{n}\else
  \'{n}\fi{}ska}}, \bibinfo{author}{\bibfnamefont{M.}~\bibnamefont{Rossi}},
  \bibinfo{author}{\bibfnamefont{A.}~\bibnamefont{Militaru}},
  \bibinfo{author}{\bibfnamefont{M.}~\bibnamefont{Frimmer}},
  \bibinfo{author}{\bibfnamefont{L.}~\bibnamefont{Novotny}},
  \bibinfo{author}{\bibfnamefont{P.}~\bibnamefont{Maurer}}, \bibnamefont{and}
  \bibinfo{author}{\bibfnamefont{O.}~\bibnamefont{Romero-Isart}},
  \bibinfo{journal}{PRX Quantum} \textbf{\bibinfo{volume}{4}},
  \bibinfo{pages}{030331} (\bibinfo{year}{2023}),
  \urlprefix\url{https://link.aps.org/doi/10.1103/PRXQuantum.4.030331}.

\bibitem[{\citenamefont{Møller et~al.}(2017)\citenamefont{Møller, Thomas,
  Vasilakis, Zeuthen, Tsaturyan, Balabas, Jensen, Schliesser, Hammerer, and
  Polzik}}]{Moller2017}
\bibinfo{author}{\bibfnamefont{C.~B.} \bibnamefont{Møller}},
  \bibinfo{author}{\bibfnamefont{R.~A.} \bibnamefont{Thomas}},
  \bibinfo{author}{\bibfnamefont{G.}~\bibnamefont{Vasilakis}},
  \bibinfo{author}{\bibfnamefont{E.}~\bibnamefont{Zeuthen}},
  \bibinfo{author}{\bibfnamefont{Y.}~\bibnamefont{Tsaturyan}},
  \bibinfo{author}{\bibfnamefont{M.}~\bibnamefont{Balabas}},
  \bibinfo{author}{\bibfnamefont{K.}~\bibnamefont{Jensen}},
  \bibinfo{author}{\bibfnamefont{A.}~\bibnamefont{Schliesser}},
  \bibinfo{author}{\bibfnamefont{K.}~\bibnamefont{Hammerer}}, \bibnamefont{and}
  \bibinfo{author}{\bibfnamefont{E.~S.} \bibnamefont{Polzik}},
  \bibinfo{journal}{Nature} \textbf{\bibinfo{volume}{547}},
  \bibinfo{pages}{191} (\bibinfo{year}{2017}), ISSN \bibinfo{issn}{1476-4687},
  \urlprefix\url{https://doi.org/10.1038/nature22980}.

\bibitem[{\citenamefont{Higginbottom et~al.}(2018)\citenamefont{Higginbottom,
  Campbell, Araneda, Fang, Colombe, Buchler, and Lam}}]{Higginbottom2018}
\bibinfo{author}{\bibfnamefont{D.~B.} \bibnamefont{Higginbottom}},
  \bibinfo{author}{\bibfnamefont{G.~T.} \bibnamefont{Campbell}},
  \bibinfo{author}{\bibfnamefont{G.}~\bibnamefont{Araneda}},
  \bibinfo{author}{\bibfnamefont{F.}~\bibnamefont{Fang}},
  \bibinfo{author}{\bibfnamefont{Y.}~\bibnamefont{Colombe}},
  \bibinfo{author}{\bibfnamefont{B.~C.} \bibnamefont{Buchler}},
  \bibnamefont{and} \bibinfo{author}{\bibfnamefont{P.~K.} \bibnamefont{Lam}},
  \bibinfo{journal}{Scientific Reports} \textbf{\bibinfo{volume}{8}},
  \bibinfo{pages}{2045} (\bibinfo{year}{2018}),
  \urlprefix\url{https://doi.org/10.1038/s41598-017-18637-8}.

\bibitem[{\citenamefont{Araneda et~al.}(2020)\citenamefont{Araneda, Cerchiari,
  Higginbottom, Holz, Lakhmanskiy, Obšil, Colombe, and Blatt}}]{araneda2020}
\bibinfo{author}{\bibfnamefont{G.}~\bibnamefont{Araneda}},
  \bibinfo{author}{\bibfnamefont{G.}~\bibnamefont{Cerchiari}},
  \bibinfo{author}{\bibfnamefont{D.~B.} \bibnamefont{Higginbottom}},
  \bibinfo{author}{\bibfnamefont{P.~C.} \bibnamefont{Holz}},
  \bibinfo{author}{\bibfnamefont{K.}~\bibnamefont{Lakhmanskiy}},
  \bibinfo{author}{\bibfnamefont{P.}~\bibnamefont{Obšil}},
  \bibinfo{author}{\bibfnamefont{Y.}~\bibnamefont{Colombe}}, \bibnamefont{and}
  \bibinfo{author}{\bibfnamefont{R.}~\bibnamefont{Blatt}},
  \bibinfo{journal}{Review of Scientific Instruments}
  \textbf{\bibinfo{volume}{91}}, \bibinfo{pages}{113201}
  (\bibinfo{year}{2020}), \urlprefix\url{https://doi.org/10.1063/5.0020661}.

\bibitem[{\citenamefont{Dania et~al.}(2022)\citenamefont{Dania, Heidegger,
  Bykov, Cerchiari, Araneda, and Northup}}]{dania2022}
\bibinfo{author}{\bibfnamefont{L.}~\bibnamefont{Dania}},
  \bibinfo{author}{\bibfnamefont{K.}~\bibnamefont{Heidegger}},
  \bibinfo{author}{\bibfnamefont{D.~S.} \bibnamefont{Bykov}},
  \bibinfo{author}{\bibfnamefont{G.}~\bibnamefont{Cerchiari}},
  \bibinfo{author}{\bibfnamefont{G.}~\bibnamefont{Araneda}}, \bibnamefont{and}
  \bibinfo{author}{\bibfnamefont{T.~E.} \bibnamefont{Northup}},
  \bibinfo{journal}{Phys. Rev. Lett.} \textbf{\bibinfo{volume}{129}},
  \bibinfo{pages}{013601} (\bibinfo{year}{2022}),
  \urlprefix\url{https://link.aps.org/doi/10.1103/PhysRevLett.129.013601}.

\bibitem[{\citenamefont{Eschner et~al.}(2001)\citenamefont{Eschner, Raab,
  Schmidt-Kaler, and Blatt}}]{Eschner2001}
\bibinfo{author}{\bibfnamefont{J.}~\bibnamefont{Eschner}},
  \bibinfo{author}{\bibfnamefont{C.}~\bibnamefont{Raab}},
  \bibinfo{author}{\bibfnamefont{F.}~\bibnamefont{Schmidt-Kaler}},
  \bibnamefont{and} \bibinfo{author}{\bibfnamefont{R.}~\bibnamefont{Blatt}},
  \bibinfo{journal}{Nature} \textbf{\bibinfo{volume}{413}},
  \bibinfo{pages}{495} (\bibinfo{year}{2001}),
  \urlprefix\url{https://doi.org/10.1038/35097017}.

\bibitem[{\citenamefont{Cerchiari
  et~al.}(2021{\natexlab{b}})\citenamefont{Cerchiari, Araneda, Podhora,
  Slodi\v{c}ka, Colombe, and Blatt}}]{cerchiari2021one}
\bibinfo{author}{\bibfnamefont{G.}~\bibnamefont{Cerchiari}},
  \bibinfo{author}{\bibfnamefont{G.}~\bibnamefont{Araneda}},
  \bibinfo{author}{\bibfnamefont{L.}~\bibnamefont{Podhora}},
  \bibinfo{author}{\bibfnamefont{L.}~\bibnamefont{Slodi\v{c}ka}},
  \bibinfo{author}{\bibfnamefont{Y.}~\bibnamefont{Colombe}}, \bibnamefont{and}
  \bibinfo{author}{\bibfnamefont{R.}~\bibnamefont{Blatt}},
  \bibinfo{journal}{Phys. Rev. Lett.} \textbf{\bibinfo{volume}{127}},
  \bibinfo{pages}{063603} (\bibinfo{year}{2021}{\natexlab{b}}),
  \urlprefix\url{https://link.aps.org/doi/10.1103/PhysRevLett.127.063603}.

\bibitem[{\citenamefont{Deli{\'c} et~al.}(2020)\citenamefont{Deli{\'c},
  Reisenbauer, Dare, Grass, Vuleti{\'c}, Kiesel, and Aspelmeyer}}]{Delic2020}
\bibinfo{author}{\bibfnamefont{U.}~\bibnamefont{Deli{\'c}}},
  \bibinfo{author}{\bibfnamefont{M.}~\bibnamefont{Reisenbauer}},
  \bibinfo{author}{\bibfnamefont{K.}~\bibnamefont{Dare}},
  \bibinfo{author}{\bibfnamefont{D.}~\bibnamefont{Grass}},
  \bibinfo{author}{\bibfnamefont{V.}~\bibnamefont{Vuleti{\'c}}},
  \bibinfo{author}{\bibfnamefont{N.}~\bibnamefont{Kiesel}}, \bibnamefont{and}
  \bibinfo{author}{\bibfnamefont{M.}~\bibnamefont{Aspelmeyer}},
  \bibinfo{journal}{Science} \textbf{\bibinfo{volume}{367}},
  \bibinfo{pages}{892} (\bibinfo{year}{2020}), ISSN \bibinfo{issn}{0036-8075},
  \urlprefix\url{https://doi.org/10.1126/science.aba3993}.

\bibitem[{\citenamefont{Martinetz et~al.}(2021)\citenamefont{Martinetz,
  Hornberger, and Stickler}}]{Martinetz_2021}
\bibinfo{author}{\bibfnamefont{L.}~\bibnamefont{Martinetz}},
  \bibinfo{author}{\bibfnamefont{K.}~\bibnamefont{Hornberger}},
  \bibnamefont{and} \bibinfo{author}{\bibfnamefont{B.~A.}
  \bibnamefont{Stickler}}, \bibinfo{journal}{New Journal of Physics}
  \textbf{\bibinfo{volume}{23}}, \bibinfo{pages}{093001}
  (\bibinfo{year}{2021}),
  \urlprefix\url{https://dx.doi.org/10.1088/1367-2630/ac1c82}.

\bibitem[{\citenamefont{van~der Laan et~al.}(2021)\citenamefont{van~der Laan,
  Tebbenjohanns, Reimann, Vijayan, Novotny, and Frimmer}}]{van2021sub}
\bibinfo{author}{\bibfnamefont{F.}~\bibnamefont{van~der Laan}},
  \bibinfo{author}{\bibfnamefont{F.}~\bibnamefont{Tebbenjohanns}},
  \bibinfo{author}{\bibfnamefont{R.}~\bibnamefont{Reimann}},
  \bibinfo{author}{\bibfnamefont{J.}~\bibnamefont{Vijayan}},
  \bibinfo{author}{\bibfnamefont{L.}~\bibnamefont{Novotny}}, \bibnamefont{and}
  \bibinfo{author}{\bibfnamefont{M.}~\bibnamefont{Frimmer}},
  \bibinfo{journal}{Phys. Rev. Lett.} \textbf{\bibinfo{volume}{127}},
  \bibinfo{pages}{123605} (\bibinfo{year}{2021}),
  \urlprefix\url{https://link.aps.org/doi/10.1103/PhysRevLett.127.123605}.

\bibitem[{\citenamefont{van~de Hulst}(1981)}]{hulst1981light}
\bibinfo{author}{\bibfnamefont{H.}~\bibnamefont{van~de Hulst}},
  \emph{\bibinfo{title}{Light Scattering by Small Particles}}, Dover Books on
  Physics (\bibinfo{publisher}{Dover Publications}, \bibinfo{year}{1981}), ISBN
  \bibinfo{isbn}{9780486642284},
  \urlprefix\url{https://books.google.at/books?id=PlHfPMVAFRcC}.

\bibitem[{\citenamefont{Bushev et~al.}(2006)\citenamefont{Bushev, Rotter,
  Wilson, Dubin, Becher, Eschner, Blatt, Steixner, Rabl, and
  Zoller}}]{Bushev2006}
\bibinfo{author}{\bibfnamefont{P.}~\bibnamefont{Bushev}},
  \bibinfo{author}{\bibfnamefont{D.}~\bibnamefont{Rotter}},
  \bibinfo{author}{\bibfnamefont{A.}~\bibnamefont{Wilson}},
  \bibinfo{author}{\bibfnamefont{F.}~\bibnamefont{Dubin}},
  \bibinfo{author}{\bibfnamefont{C.}~\bibnamefont{Becher}},
  \bibinfo{author}{\bibfnamefont{J.}~\bibnamefont{Eschner}},
  \bibinfo{author}{\bibfnamefont{R.}~\bibnamefont{Blatt}},
  \bibinfo{author}{\bibfnamefont{V.}~\bibnamefont{Steixner}},
  \bibinfo{author}{\bibfnamefont{P.}~\bibnamefont{Rabl}}, \bibnamefont{and}
  \bibinfo{author}{\bibfnamefont{P.}~\bibnamefont{Zoller}},
  \bibinfo{journal}{Phys. Rev. Lett.} \textbf{\bibinfo{volume}{96}},
  \bibinfo{pages}{043003} (\bibinfo{year}{2006}),
  \urlprefix\url{https://doi.org/10.1103/PhysRevLett.96.043003}.

\bibitem[{\citenamefont{Clerk et~al.}(2010)\citenamefont{Clerk, Devoret,
  Girvin, Marquardt, and Schoelkopf}}]{Clerk}
\bibinfo{author}{\bibfnamefont{A.~A.} \bibnamefont{Clerk}},
  \bibinfo{author}{\bibfnamefont{M.~H.} \bibnamefont{Devoret}},
  \bibinfo{author}{\bibfnamefont{S.~M.} \bibnamefont{Girvin}},
  \bibinfo{author}{\bibfnamefont{F.}~\bibnamefont{Marquardt}},
  \bibnamefont{and} \bibinfo{author}{\bibfnamefont{R.~J.}
  \bibnamefont{Schoelkopf}}, \bibinfo{journal}{Rev. Mod. Phys.}
  \textbf{\bibinfo{volume}{82}}, \bibinfo{pages}{1155} (\bibinfo{year}{2010}),
  \urlprefix\url{https://doi.org/10.1103/RevModPhys.82.1155}.

\bibitem[{\citenamefont{Romero-Isart et~al.}(2011)\citenamefont{Romero-Isart,
  Pflanzer, Blaser, Kaltenbaek, Kiesel, Aspelmeyer, and
  Cirac}}]{RomeroIsart2011}
\bibinfo{author}{\bibfnamefont{O.}~\bibnamefont{Romero-Isart}},
  \bibinfo{author}{\bibfnamefont{A.~C.} \bibnamefont{Pflanzer}},
  \bibinfo{author}{\bibfnamefont{F.}~\bibnamefont{Blaser}},
  \bibinfo{author}{\bibfnamefont{R.}~\bibnamefont{Kaltenbaek}},
  \bibinfo{author}{\bibfnamefont{N.}~\bibnamefont{Kiesel}},
  \bibinfo{author}{\bibfnamefont{M.}~\bibnamefont{Aspelmeyer}},
  \bibnamefont{and} \bibinfo{author}{\bibfnamefont{J.~I.} \bibnamefont{Cirac}},
  \bibinfo{journal}{Phys. Rev. Lett.} \textbf{\bibinfo{volume}{107}},
  \bibinfo{pages}{020405} (\bibinfo{year}{2011}),
  \urlprefix\url{https://link.aps.org/doi/10.1103/PhysRevLett.107.020405}.

\bibitem[{\citenamefont{Brawley et~al.}(2016)\citenamefont{Brawley, Vanner,
  Larsen, Schmid, Boisen, and Bowen}}]{Brawley2016}
\bibinfo{author}{\bibfnamefont{G.~A.} \bibnamefont{Brawley}},
  \bibinfo{author}{\bibfnamefont{M.~R.} \bibnamefont{Vanner}},
  \bibinfo{author}{\bibfnamefont{P.~E.} \bibnamefont{Larsen}},
  \bibinfo{author}{\bibfnamefont{S.}~\bibnamefont{Schmid}},
  \bibinfo{author}{\bibfnamefont{A.}~\bibnamefont{Boisen}}, \bibnamefont{and}
  \bibinfo{author}{\bibfnamefont{W.~P.} \bibnamefont{Bowen}},
  \bibinfo{journal}{Nature Communications} \textbf{\bibinfo{volume}{7}},
  \bibinfo{pages}{10988} (\bibinfo{year}{2016}), ISSN
  \bibinfo{issn}{2041-1723},
  \urlprefix\url{https://doi.org/10.1038/ncomms10988}.

\bibitem[{\citenamefont{Aspelmeyer et~al.}(2014)\citenamefont{Aspelmeyer,
  Kippenberg, and Marquardt}}]{Aspelmeyer2014}
\bibinfo{author}{\bibfnamefont{M.}~\bibnamefont{Aspelmeyer}},
  \bibinfo{author}{\bibfnamefont{T.~J.} \bibnamefont{Kippenberg}},
  \bibnamefont{and}
  \bibinfo{author}{\bibfnamefont{F.}~\bibnamefont{Marquardt}},
  \bibinfo{journal}{Rev. Mod. Phys.} \textbf{\bibinfo{volume}{86}},
  \bibinfo{pages}{1391} (\bibinfo{year}{2014}),
  \urlprefix\url{https://doi.org/10.1103/RevModPhys.86.1391}.

\bibitem[{\citenamefont{Hoang et~al.}(2016)\citenamefont{Hoang, Ma, Ahn, Bang,
  Robicheaux, Yin, and Li}}]{Hoang2016}
\bibinfo{author}{\bibfnamefont{T.~M.} \bibnamefont{Hoang}},
  \bibinfo{author}{\bibfnamefont{Y.}~\bibnamefont{Ma}},
  \bibinfo{author}{\bibfnamefont{J.}~\bibnamefont{Ahn}},
  \bibinfo{author}{\bibfnamefont{J.}~\bibnamefont{Bang}},
  \bibinfo{author}{\bibfnamefont{F.}~\bibnamefont{Robicheaux}},
  \bibinfo{author}{\bibfnamefont{Z.-Q.} \bibnamefont{Yin}}, \bibnamefont{and}
  \bibinfo{author}{\bibfnamefont{T.}~\bibnamefont{Li}}, \bibinfo{journal}{Phys.
  Rev. Lett.} \textbf{\bibinfo{volume}{117}}, \bibinfo{pages}{123604}
  (\bibinfo{year}{2016}),
  \urlprefix\url{https://link.aps.org/doi/10.1103/PhysRevLett.117.123604}.

\bibitem[{\citenamefont{Ahn et~al.}(2018)\citenamefont{Ahn, Xu, Bang, Deng,
  Hoang, Han, Ma, and Li}}]{Ahn2018}
\bibinfo{author}{\bibfnamefont{J.}~\bibnamefont{Ahn}},
  \bibinfo{author}{\bibfnamefont{Z.}~\bibnamefont{Xu}},
  \bibinfo{author}{\bibfnamefont{J.}~\bibnamefont{Bang}},
  \bibinfo{author}{\bibfnamefont{Y.-H.} \bibnamefont{Deng}},
  \bibinfo{author}{\bibfnamefont{T.~M.} \bibnamefont{Hoang}},
  \bibinfo{author}{\bibfnamefont{Q.}~\bibnamefont{Han}},
  \bibinfo{author}{\bibfnamefont{R.-M.} \bibnamefont{Ma}}, \bibnamefont{and}
  \bibinfo{author}{\bibfnamefont{T.}~\bibnamefont{Li}}, \bibinfo{journal}{Phys.
  Rev. Lett.} \textbf{\bibinfo{volume}{121}}, \bibinfo{pages}{033603}
  (\bibinfo{year}{2018}),
  \urlprefix\url{https://link.aps.org/doi/10.1103/PhysRevLett.121.033603}.

\bibitem[{\citenamefont{Kuhn et~al.}(2017)\citenamefont{Kuhn, Kosloff,
  Stickler, Patolsky, Hornberger, Arndt, and Millen}}]{Kuhn2017}
\bibinfo{author}{\bibfnamefont{S.}~\bibnamefont{Kuhn}},
  \bibinfo{author}{\bibfnamefont{A.}~\bibnamefont{Kosloff}},
  \bibinfo{author}{\bibfnamefont{B.~A.} \bibnamefont{Stickler}},
  \bibinfo{author}{\bibfnamefont{F.}~\bibnamefont{Patolsky}},
  \bibinfo{author}{\bibfnamefont{K.}~\bibnamefont{Hornberger}},
  \bibinfo{author}{\bibfnamefont{M.}~\bibnamefont{Arndt}}, \bibnamefont{and}
  \bibinfo{author}{\bibfnamefont{J.}~\bibnamefont{Millen}},
  \bibinfo{journal}{Optica} \textbf{\bibinfo{volume}{4}}, \bibinfo{pages}{356}
  (\bibinfo{year}{2017}),
  \urlprefix\url{https://opg.optica.org/optica/abstract.cfm?URI=optica-4-3-356}.

\bibitem[{\citenamefont{Krutyanskiy et~al.}(2023)\citenamefont{Krutyanskiy,
  Galli, Krcmarsky, Baier, Fioretto, Pu, Mazloom, Sekatski, Canteri, Teller
  et~al.}}]{Krutyanskiy2023}
\bibinfo{author}{\bibfnamefont{V.}~\bibnamefont{Krutyanskiy}},
  \bibinfo{author}{\bibfnamefont{M.}~\bibnamefont{Galli}},
  \bibinfo{author}{\bibfnamefont{V.}~\bibnamefont{Krcmarsky}},
  \bibinfo{author}{\bibfnamefont{S.}~\bibnamefont{Baier}},
  \bibinfo{author}{\bibfnamefont{D.}~\bibnamefont{Fioretto}},
  \bibinfo{author}{\bibfnamefont{Y.}~\bibnamefont{Pu}},
  \bibinfo{author}{\bibfnamefont{A.}~\bibnamefont{Mazloom}},
  \bibinfo{author}{\bibfnamefont{P.}~\bibnamefont{Sekatski}},
  \bibinfo{author}{\bibfnamefont{M.}~\bibnamefont{Canteri}},
  \bibinfo{author}{\bibfnamefont{M.}~\bibnamefont{Teller}},
  \bibnamefont{et~al.}, \bibinfo{journal}{Physical Review Letters}
  \textbf{\bibinfo{volume}{130}}, \bibinfo{pages}{050803}
  (\bibinfo{year}{2023}),
  \urlprefix\url{https://link.aps.org/doi/10.1103/PhysRevLett.130.050803}.

\bibitem[{\citenamefont{Cerchiari et~al.}(2024)\citenamefont{Cerchiari, Weiser,
  Faorlin, Panzl, and Lafenthaler}}]{Cerchiari2024}
\bibinfo{author}{\bibfnamefont{G.}~\bibnamefont{Cerchiari}},
  \bibinfo{author}{\bibfnamefont{Y.}~\bibnamefont{Weiser}},
  \bibinfo{author}{\bibfnamefont{T.}~\bibnamefont{Faorlin}},
  \bibinfo{author}{\bibfnamefont{L.}~\bibnamefont{Panzl}}, \bibnamefont{and}
  \bibinfo{author}{\bibfnamefont{T.}~\bibnamefont{Lafenthaler}}
  (\bibinfo{year}{2024}),
  \urlprefix\url{https://doi.org/10.5281/zenodo.10656593}.

\bibitem[{\citenamefont{Novotny and Hecht}(2012)}]{Novotny2012book}
\bibinfo{author}{\bibfnamefont{L.}~\bibnamefont{Novotny}} \bibnamefont{and}
  \bibinfo{author}{\bibfnamefont{B.}~\bibnamefont{Hecht}},
  \emph{\bibinfo{title}{Principles of Nano-Optics}}
  (\bibinfo{publisher}{Cambridge University Press}, \bibinfo{year}{2012}), ISBN
  \bibinfo{isbn}{978-1-107-00546-4},
  \urlprefix\url{www.cambridge.org/9781107005464}.

\end{thebibliography}

\hfill \break

\renewcommand{\thefigure}{A.\arabic{figure}}
\setcounter{figure}{0}
\renewcommand{\theequation}{A.\arabic{equation}}
\setcounter{equation}{0}
\section{Appendix A: Mirror's transfer function}\label{sec:image_position}

In this section, we derive the expression Eq.~\eqref{eq:dP_full}. This calculation is not discussed in the main text since the presence of a hemispherical mirror introduces only minor modifications to the calculations for emitters in free space, which can instead be readily found in textbooks such as Ref.~\cite{Novotny2012book}. 

The vector potential $\bm{A}$ under the Lorentz gauge condition for a free-space emitter is~\cite{Novotny2012book}
\begin{equation}
    \label{eq:vector_potential}
    \bm{A}\left(\bm{x}'\right)=\frac{\hat{\bm{p}}}{4 \pi \epsilon_0 c^2}\int j\left(\bm{x}\right) g_m\left(\hat{\bm{n}},\bm{x}\right) d^3\bm{x},
\end{equation}
where we assumed that the current $\bm{j}$ is composed of charges moving along the direction of $\hat{\bm{p}}$. This is the case with linear dipolar scatterers in which the induced dipole is aligned with the polarization of the illuminating field.
The function $g_m\left(\bm{x}',\bm{x}\right)$ corresponds to the integral Green's function of the Helmholtz equation and has the expression
\begin{equation}
    g_m\left(\bm{x}',\bm{x}\right) = \frac{e^{i \lvert\bm{k}\rvert \lvert \bm{x}'- \bm{x} \rvert}}{ \lvert\bm{x}'-\bm{x}\rvert} \approx \frac{e^{i \lvert\bm{k}\rvert \lvert \bm{x}'\rvert}}{ \lvert\bm{x}'\rvert} \exp{\left(-i\lvert \bm{k}\rvert\hat{\bm{n}}\cdot\bm{x}\right)} \, ,
\end{equation}
where in the last step we used the far-field approximation. The Green's function is modified by the presence of the mirror because the latter generates an image of a point-like emitter with the expression~\cite{cerchiari2021dipole}
\begin{equation}
g_i\left(\hat{\bm{n}},\bm{x}\right) = -\sqrt{R}\frac{e^{i \lvert\bm{k}\rvert \lvert \bm{x}'\rvert}}{ \lvert\bm{x}'\rvert}\exp{\left(i\lvert \bm{k}\rvert\left(\hat{\bm{n}}\cdot\bm{x}-2R_s\right)\right)}  \; ,
\end{equation}
where $\sqrt{R}$ is the reflection coefficient for the electric field of the mirror, $R_s$ is the radius of the mirror and $\hat{\bm{n}}$ satisfies the equation $\bm{x}'=\lvert \bm{x}' \rvert \hat{\bm{n}}$. Compared to the expression for $g_m$, here the sign of $\bm{x}$ is reversed because the image is formed on the opposite side of the mirror's center of curvature. The image interferes with the primary field emitted by the scatterer, thus the function $g_m$ in Eq.~\eqref{eq:vector_potential} has to be substituted with the Green's function $g_c=g_m+g_i$ of the compound system. We approximate the mirror so that it has perfect reflectivity ($\sqrt{R}=1$). Under this condition, we obtain the following expression for $g_c$
\begin{equation}
    g_c\left(\hat{\bm{n}},\bm{x}\right)
    = -2i \frac{e^{i \lvert\bm{k}\rvert \lvert \bm{x}'\rvert}}{ \lvert\bm{x}'\rvert} \exp{\left(-iR_s\right)}\left(\sin{\left(\lvert \bm{k}\rvert\hat{\bm{n}}\cdot\bm{x}-R_s\right)}\right)  \; .
\end{equation}

In order to suppress the emission of any vector potential produced by a dipolar point scatterer located in the center of curvature, $g_c\left(\hat{\bm{n}},\bm{0}\right)=0$ must hold. This equation is satisfied for $R_s = \pi m$ with $m$ being an integer number. Under this condition, we find
\begin{equation}
    g_c\left(\hat{\bm{n}},\bm{x}\right)= \mp 2i \frac{e^{i \lvert\bm{k}\rvert \lvert \bm{x}'\rvert}}{ \lvert\bm{x}'\rvert} \sin{\left(\lvert \bm{k}\rvert\hat{\bm{n}}\cdot\bm{x}\right)}  \; .
\end{equation}
These solutions are equivalent for determining the intensity at the detector and we have opted for the positive sign in the main text.

The electric field $\bm{E}$ can be calculated from the vector potential via the expression~\cite{Novotny2012book}
\begin{equation}
    \bm{E}\left(\bm{x}'\right) = i  c \lvert \bm{k}\rvert \left(I + \frac{1}{\lvert \bm{k}\rvert^2} \nabla' \nabla' \right) \bm{A}\left(\bm{x}'\right) \;.
\end{equation}

We performed this calculation in Ref.~\cite{Cerchiari2024} and we applied the far-field approximation that consists in retaining only the leading term in $1/\lvert \bm{x}' \rvert$. Under this approximation, we find
\begin{equation}
    \label{eq:electric_field}
    \bm{E}\left(\hat{\bm{n}}\right) = \left(G\left(\hat{\bm{n}}\right)\cdot\hat{\bm{p}}\right)\int 
    j\left(\bm{x}\right) g\left(\hat{\bm{n}},\bm{x}\right) d^3\bm{x} \; ,
\end{equation}
where $G\left(\hat{\bm{n}}\right)$ has the expression 
\begin{equation}
    \label{eq:Green_matrix}
    G\left(\hat{\bm{n}}\right) = - \frac{\lvert \bm{k}\rvert}{4 \pi \epsilon_0 c}  \left(\begin{matrix}
1-c^2_\theta c^2_\phi & -s^2_\theta c_\phi s_\phi & -c_\theta s_\theta c_\phi\\
-s^2_\theta c_\phi s_\phi  & 1-s^2_\theta s^2_\phi & -c_\theta s_\theta s_\phi\\
-c_\theta s_\theta c_\phi & -c_\theta s_\theta s_\phi & s^2_\theta
\end{matrix} \right)\; .
\end{equation}
In the formula we used the abbreviations $c_\alpha = \cos{\left(\alpha\right)}$ and $s_\alpha = \sin{\left(\alpha\right)}$. This result is valid for both $g_m$ and $g_c$ as initial Green's function because they both depend on the position $\bm{x}'$ via the same function at the prefactor. 

The differential power emitted $d\tilde{P}$ emitted in the solid angle $d\Omega$ can be calculated from the electric field via the formula
\begin{equation}
    \frac{d\tilde{P}}{d\Omega}=\frac{\epsilon_0 c}{2}\lvert\bm{E}\rvert^2\; .
\end{equation}
The square of the electric field Eq.~\eqref{eq:electric_field} has two contributions. One is
\begin{equation}
    \lvert G\left(\hat{\bm{n}}\right)\cdot\hat{\bm{p}}\rvert^2 = \frac{\lvert \bm{k}\rvert^2}{16 \pi^2 \epsilon_0^2 c^2}\left(1-\left(\hat{\bm{n}}\cdot \hat{\bm{p}}\right)^2\right),
\end{equation}
which imprints the dipolar pattern of emission and the other is the square of the integral part. The final formula for the differential power emitted per unit solid angle is
\begin{equation}
    \frac{d\tilde{P}}{d\Omega}=\frac{\lvert \bm{k}\rvert^2}{32 \pi^2 \epsilon_0 c}\left(1-\left(\hat{\bm{n}}\cdot \hat{\bm{p}}\right)^2\right) \left\lvert\int j\left(\bm{x}\right) g\left(\hat{\bm{n}},\bm{x}\right) d^3\bm{x}\right\rvert^2\;
\end{equation}
and can be integrated over the measurement and control region to obtain the total radiated power.
 
For a dipole radiation field induced by a plane wave excitation, the expression for the current is
\begin{equation}
    \bm{j}\left(\bm{x}\right)=i c \lvert \bm{k}\rvert \rho\left(\bm{x}\right) \hat{\bm{p}} \exp{\left(i \bm{k}\cdot\bm{x}\right)}\, ,
\end{equation}
where $\rho\left(\bm{x}\right)$ is the dipole moment density induced by the illuminating beam with polarization $\hat{\bm{p}}$, $c \lvert \bm{k} \rvert$ corresponds to the angular frequency of the illuminating radiation and $\exp{\left(i \bm{k}\cdot\bm{x}\right)}$ is the phase accumulated by the illuminating wave by traveling inside the object. For a point-like dipole, the density is $\rho\left(\bm{x}\right)=p\delta\left(\bm{x}\right)$, which leads in free space to a differential emitted power (Eq.~\eqref{eq:dP_dipole}) of
\begin{equation}
    \frac{dP}{d\Omega} = p^2 \frac{c\lvert \bm{k}\rvert^4}{32 \pi^2 \epsilon_0}\left(1-\left(\hat{\bm{n}}\cdot \hat{\bm{p}}\right)^2\right)\; ,
\end{equation}
with a corresponding total emitted power $P_{dip}$
\begin{equation}
    P_{dip}=p^2\frac{c\lvert \bm{k}\rvert^4 }{12 \pi \epsilon_0}\;.
\end{equation}

Finally, in the main text, we assume the measurement to take place on the unitary sphere ($\lvert\bm{x}'\rvert=1$) and disregard constant phase terms, which are not useful for an evaluation of the radiated power ($e^{i \lvert\bm{k}\rvert}=1$). With the new definition of Green's functions, we can re-write the differential radiated power $d\tilde{P}$ as
\begin{equation}
    \frac{d\tilde{P}}{d\Omega}= \frac{1}{p^2}\left\lvert\int\rho\left(\bm{x}\right) g\left(\hat{\bm{n}},\bm{x}\right)e^{i \bm{k}\cdot\bm{x}} d^3\bm{x}\right\rvert^2 \frac{dP}{d\Omega}\;,
\end{equation}
which is Eq.~\eqref{eq:dP_full} of the main text.

\renewcommand{\thefigure}{B.\arabic{figure}}
\setcounter{figure}{0}
\renewcommand{\theequation}{B.\arabic{equation}}
\setcounter{equation}{0}
\section{Appendix B: Radiated power from a sphere}\label{sec:sphere_power}

In the case of a single atom emitting visible fluorescence radiation, the dipole distribution $\rho\left(x\right)$ can be considered a delta function and the suppression of the scattering rate is complete. This approximation breaks down for scatterers whose spatial extension becomes comparable to the wavelength of the scattered field and we discuss in this appendix this limitation considering as a scatterer a levitated sphere of radius $R_0$. The dipole density function is $\rho\left(r\right)=\rho H_\theta\left(r-R_0\right)$, where $H_\theta$ is the Heaviside function. Since we wish to evaluate the contribution to scattering for a scatterer whose size is comparable to the wavelength, we will provide an estimate regarding the error introduced by the approximation of Eqs.~\eqref{eq:green_free_leading} and \eqref{eq:green_mirror_leading} by presenting also the calculations in which $g_m$ and $g_c$ are expanded to second order in $\lvert\bm{x}\rvert/\lambda$, i.e. by using the expressions 
\begin{align}
    e^{i \bm{k}\cdot \bm{x}} g_m\left(\hat{\bm{n}},\bm{x}\right) &\approx 1 + i \lvert \bm{k}\rvert \left(\hat{\bm{k}}\cdot\bm{x}\ + \hat{\bm{n}}\cdot\bm{x}\right)\; +\\
    & - \frac{\lvert \bm{k}\rvert^2}{2}\left(\hat{\bm{k}}\cdot\bm{x} + \hat{\bm{n}}\cdot\bm{x}\right)^2, \nonumber \\
    e^{i \bm{k}\cdot \bm{x}} 
    g_c\left(\hat{\bm{n}},\bm{x}\right) &\approx 2 \lvert \bm{k}\rvert \left(\hat{\bm{n}}\cdot\bm{x}\right) + i \lvert \bm{k}\rvert^2 \left(\hat{\bm{k}}\cdot\bm{x}\right)\left( \hat{\bm{n}}\cdot\bm{x}\right)   \; .
\end{align}
 Explicitly, the power emitted in the measurement region $P_m$ and in the control region $P_c$ is given by
\begin{align}
\label{eq:power_measurement_region}
    P_m\left(\textrm{NA}\right)&=2\int_0^{\arcsin{\left(\textrm{NA}\right)}}d\theta\sin{\left(\theta\right)}\int_0^{2\pi}d\phi \frac{dP_m}{d\Omega}\;,\\
    P_c\left(\textrm{NA}\right)&=\int_{\arcsin{\left(\textrm{NA}\right)}}^{\pi/2}d\theta\sin{\left(\theta\right)}\int_0^{2\pi}d\phi \frac{dP_c}{d\Omega} \; .
\end{align}
The factor of two in Eq.~\eqref{eq:power_measurement_region} accounts for both sections of the measurement region.

\begin{figure}
\includegraphics[width=\columnwidth]{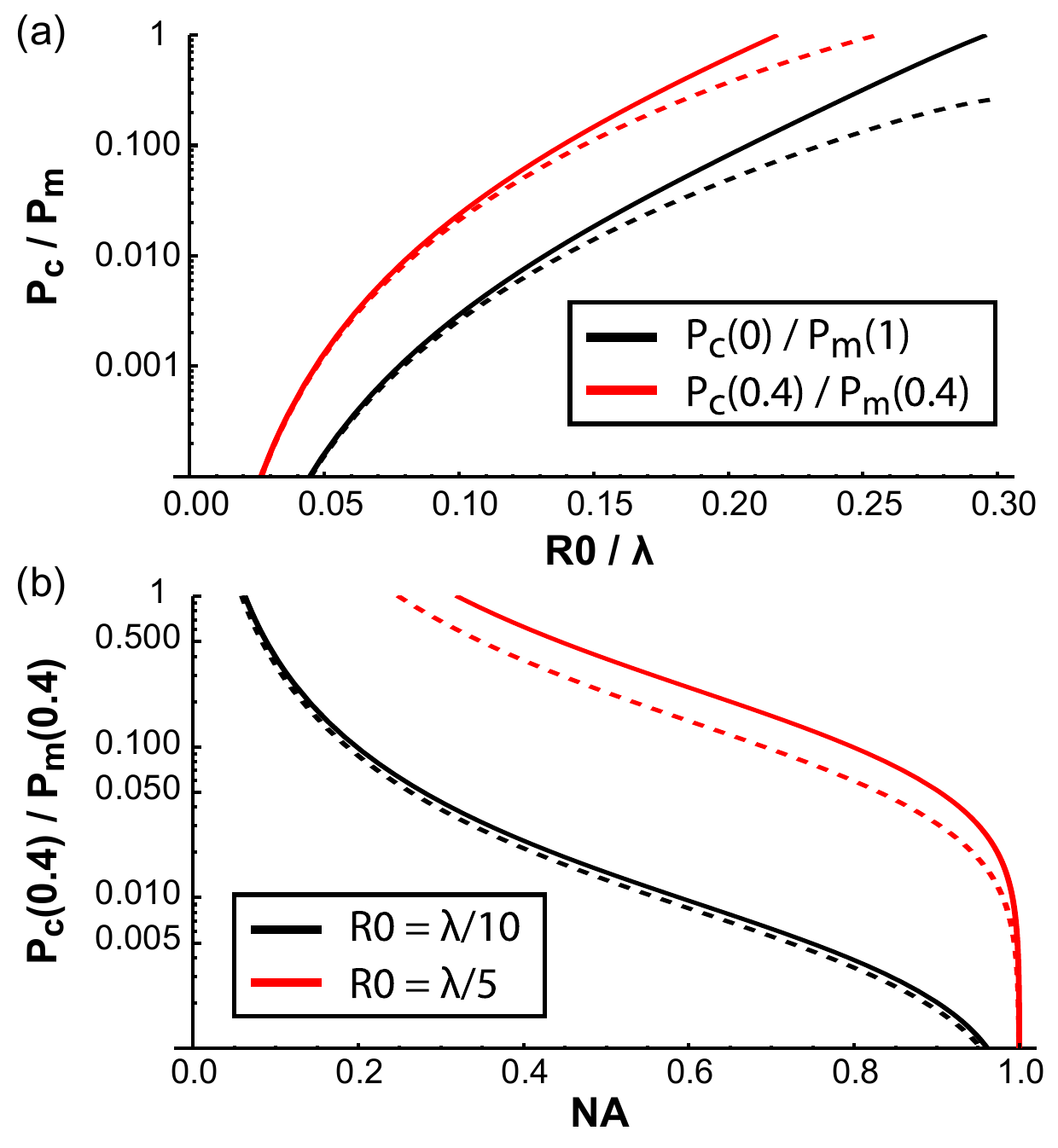}
\caption{Scattering rate of a levitated sphere. (a) Emitted power as a function of the sphere's radius $R_0$. (b) Ratio as a function of the numerical aperture $\textrm{NA}$ of the measurement region. The solid line corresponds to the calculation up to the second order in $R0/\lambda$ and the dashed line up to the fourth order.}
\label{fig:PcPm}
\end{figure}

We quantify the control of the emission by computing the ratio $P_c/P_m$. For instance, the ratio $P_c(0)/P_m(1)$ corresponds to the power emitted in the presence of a full hemispherical mirror without a hole, divided by the power emitted when the mirror is removed from the system. This ratio quantifies how much the emission of the levitated sphere can be suppressed via the mirror. This quantity can only be measured in a sequential experiment in which the levitated sphere is moved in and out from the center of curvature of such a mirror and it is here shown for reference. Relevant for the discussion in the main text is the ratio $P_c(\textrm{NA})/P_m(\textrm{NA})$, which is the power emitted in the control region divided by the power emitted in the measurement region if the hemispherical hollow mirror is controlling the emission. The two quantities ($P_c(0)/P_m(1)$ and $P_c(0.4)/P_m(0.4)$) are presented in Fig.~\ref{fig:PcPm}(a) as a function of the radius of the levitated sphere. We selected a numerical aperture of 0.4 because it has been already demonstrated to be sufficient for optomechanical measurements at the quantum level~\cite{cerchiari2021one}. We see that, although the numerical aperture adhering to the condition $\textrm{NA}=0.4$ corresponds to the measurement region being 8\% of the total solid angle, the amount of radiation exiting in the control region is lower than the one emitted in the measurement region up to about $R_0=\lambda/4$. For comparison, in free space, the radiation power scattered outside the measurement region is seven times higher than in the measurement region itself. Figure.~\ref{fig:PcPm}(b) presents the $P_c(\textrm{NA})/P_m(\textrm{NA})$ as a function of the numerical aperture of the mirror's hole $\textrm{NA}$ for a sphere of radius $R_0=\lambda/10$ and a sphere of radius $R_0=\lambda/5$. We see that, the approximation is validated in epxerimental conditions which are used in laboratories such as described in Ref.~\cite{dania2022}. 

\renewcommand{\thefigure}{C.\arabic{figure}}
\setcounter{figure}{0}
\renewcommand{\theequation}{C.\arabic{equation}}
\setcounter{equation}{0}
\section{Appendix C: Back action suppression}\label{sec:variance}

In the regime in which shot noise dominates the fluctuations, the power spectral density of the radiated power is~\cite{Tebbenjohanns2019, cerchiari2021dipole}
\begin{equation}
    ds_{pp} = \frac{\hbar \lvert \bm{k}\rvert c}{2 \pi} dP \;,
\end{equation}
which induces fluctuations in the radiation pressure force $ds_{ba}$ along the direction $\hat{\bm{x}}_0$ equal to
\begin{equation}
    ds_{ba} = \frac{1}{c^2}\left(\hat{\bm{n}}\cdot \bm{x}_0\right)^2 ds_{pp}\;.
\end{equation}
The integral of this expression (see Eq.~\eqref{eq:total_back_action}) over the full solid angle gives
\begin{equation}
    S_{ba}^z = S_{ba}\left(\hat{\bm{z}}, \pi/2\right) = \frac{2 \hbar \lvert \bm{k}\rvert P_{dip}}{10 \pi c}
\end{equation}
and for the three directions of space
\begin{align}
    \frac{S_{ba}\left(\hat{\bm{x}}, \theta\right)}{S_{ba}^z} &= \frac{\sin^2{\left(\theta/2\right)}}{32} \left(100 + 95 c_1+36 c_1+9 c_3\right) \;,\\
    \frac{S_{ba}\left(\hat{\bm{y}}, \theta\right)}{S_{ba}^z}  &= \frac{\sin^2{\left(\theta/2\right)}}{32}\left(140 + 85 c_1+12 c_1+3 c_3\right) \;,\\
    \frac{S_{ba}\left(\hat{\bm{z}}, \theta\right)}{S_{ba}^z}  &= 1-\frac{1}{16} c_1 \left(13+3 c_2\right) \;,
\end{align}
with the abbreviations $c_1=\cos{\left(\theta\right)}$, $c_2=\cos{\left(2\theta\right)}$ and $c_3=\cos{\left(3\theta\right)}$.

\end{document}